\documentstyle[12pt]{article}

\def\hybrid{\topmargin -20pt	\oddsidemargin 0pt
	\headheight 0pt	\headsep 0pt
	\textwidth 6.25in	
	\textheight 9.5in	
	\marginparwidth .875in
	\parskip 5pt plus 1pt	\jot = 1.5ex}

\hybrid

\def\baselinestretch{1.2}

\catcode`\@=11

\def\marginnote#1{}
\newcount\hour
\newcount\minute
\newtoks\amorpm
\hour=\time\divide\hour by60
\minute=\time{\multiply\hour by60 \global\advance\minute by-\hour}
\edef\standardtime{{\ifnum\hour<12 \global\amorpm={am}%
	\else\global\amorpm={pm}\advance\hour by-12 \fi
	\ifnum\hour=0 \hour=12 \fi
	\number\hour:\ifnum\minute<10 0\fi\number\minute\the\amorpm}}
\edef\militarytime{\number\hour:\ifnum\minute<10 0\fi\number\minute}

\def\draftlabel#1{{\@bsphack\if@filesw {\let\thepage\relax
   \xdef\@gtempa{\write\@auxout{\string
      \newlabel{#1}{{\@currentlabel}{\thepage}}}}}\@gtempa
   \if@nobreak \ifvmode\nobreak\fi\fi\fi\@esphack}
	\gdef\@eqnlabel{#1}}
\def\@eqnlabel{}
\def\@vacuum{}
\def\draftmarginnote#1{\marginpar{\raggedright\scriptsize\tt#1}}

\def\draft{\oddsidemargin -.5truein
	\def\@oddfoot{\sl preliminary draft \hfil
	\rm\thepage\hfil\sl\today\quad\militarytime}
	\let\@evenfoot\@oddfoot	\overfullrule 3pt
	\let\label=\draftlabel
	\let\marginnote=\draftmarginnote
   \def\@eqnnum{(\theequation)\rlap{\kern\marginparsep\tt\@eqnlabel}%
\global\let\@eqnlabel\@vacuum}  }

\def\preprint{\twocolumn\sloppy\flushbottom\parindent 2em
	\leftmargini 2em\leftmarginv .5em\leftmarginvi .5em
	\oddsidemargin -.5in	\evensidemargin -.5in
	\columnsep .4in	\footheight 0pt
	\textwidth 10.in	\topmargin  -.4in
	\headheight 12pt \topskip .4in
	\textheight 6.9in \footskip 0pt
	\def\@oddhead{\thepage\hfil\addtocounter{page}{1}\thepage}
	\let\@evenhead\@oddhead	\def\@oddfoot{}	\def\@evenfoot{} }

\def\numberbysection{\@addtoreset{equation}{section}
	\def\theequation{\thesection.\arabic{equation}}}

\def\underline#1{\relax\ifmmode\@@underline#1\else
	$\@@underline{\hbox{#1}}$\relax\fi}

\def\titlepage{\@restonecolfalse\if@twocolumn\@restonecoltrue\onecolumn
     \else \newpage \fi \thispagestyle{empty}\c@page\z@
	\def\thefootnote{\fnsymbol{footnote}} }

\def\endtitlepage{\if@restonecol\twocolumn \else \newpage \fi
	\def\thefootnote{\arabic{footnote}}
	\setcounter{footnote}{0}}  

\catcode`@=12
\relax

\def\figcap{\section*{Figure Captions\markboth
	{FIGURECAPTIONS}{FIGURECAPTIONS}}\list
	{Figure \arabic{enumi}:\hfill}{\settowidth\labelwidth{Figure
999:}
	\leftmargin\labelwidth
	\advance\leftmargin\labelsep\usecounter{enumi}}}
 \relax
\def\tablecap{\section*{Table Captions\markboth
	{TABLECAPTIONS}{TABLECAPTIONS}}\list
	{Table \arabic{enumi}:\hfill}{\settowidth\labelwidth{Table
999:}
	\leftmargin\labelwidth
	\advance\leftmargin\labelsep\usecounter{enumi}}}
 \relax
\def\reflist{\section*{References\markboth
	{REFLIST}{REFLIST}}\list
	{[\arabic{enumi}]\hfill}{\settowidth\labelwidth{[999]}
	\leftmargin\labelwidth
	\advance\leftmargin\labelsep\usecounter{enumi}}}
 \relax
\makeatletter
\newcounter{pubctr}
\def\publist{\@ifnextchar[{\@publist}{\@@publist}}
\def\@publist[#1]{\list
	{[\arabic{pubctr}]\hfill}{\settowidth\labelwidth{[999]}
	\leftmargin\labelwidth
	\advance\leftmargin\labelsep
	\@nmbrlisttrue\def\@listctr{pubctr}
	\setcounter{pubctr}{#1}\addtocounter{pubctr}{-1}}}
\def\@@publist{\list
	{[\arabic{pubctr}]\hfill}{\settowidth\labelwidth{[999]}
	\leftmargin\labelwidth
	\advance\leftmargin\labelsep
	\@nmbrlisttrue\def\@listctr{pubctr}}}
 \relax
\makeatother
\newskip\humongous \humongous=0pt plus 1000pt minus 1000pt

\newif\ifdtup

\relax

\def\thefootnote{\fnsymbol{footnote}}
\def\be{\begin{equation}}
\def\ee{\end{equation}}
\def\ba{\begin{eqnarray}}
\def\ea{\end{eqnarray}}

\begin{document}
\renewcommand{\theequation}{\arabic{equation}}
\newcommand{\beq}{\begin{equation}}
\newcommand{\eeq}[1]{\label{#1}\end{equation}}
\newcommand{\ber}{\begin{eqnarray}}
\newcommand{\eer}[1]{\label{#1}\end{eqnarray}}
\begin{titlepage}
\begin{center}

\hfill CERN--TH/95--322\\
\hfill ENSLAPP--A--566/95\\
\hfill THU--95/34\\
\hfill hep--th/9601087\\

\vskip .3in

{\large \bf TODA FIELDS OF SO(3) HYPER--KAHLER METRICS}

\vskip 0.4in

{\bf Ioannis Bakas}
\footnote{Permanent address: Department of Physics, University of
Patras,
26110 Patras, Greece}
\footnote{e--mail address: BAKAS@SURYA11.CERN.CH,
BAKAS@LAPPHP8.IN2P3.FR}\\
\vskip .1in

{\em Theory Division, CERN, 1211 Geneva 23, Switzerland, and\\
Laboratoire de Physique Theorique ENSLAPP, 74941 Annecy-le-Vieux,
France}
\footnote{Present address}\\

\vskip .3in

{\bf Konstadinos Sfetsos}
\footnote{e--mail address: SFETSOS@FYS.RUU.NL}\\
\vskip .1in

{\em Institute for Theoretical Physics, Utrecht University\\
     Princetonplein 5, TA 3508, The Netherlands}\\

\vskip .1in

\end{center}

\vskip .5in

\begin{center} {\bf ABSTRACT } \end{center}
\begin{quotation}\noindent
We examine the Toda frame formulation of the $SO(3)$--invariant
hyper--Kahler 4--metrics, namely Eguchi--Hanson, Taub--NUT and
Atiyah--Hitchin. Our method exploits the presence of a rotational
$SO(2)$ isometry, leading to the explicit construction of all
three complex structures as a singlet plus a doublet. The
Atiyah--Hitchin metric on the moduli space of BPS $SU(2)$ monopoles
with magnetic charge 2 is purely rotational.

\vskip .4in

\noindent
{\em Contribution to the proceedings of the 29th International
Symposium Ahrenshoop on the Theory of Elementary Particles, Buckow,
Germany,
29 August - 2 September 1995 (invited talk given by I. Bakas);
to appear in Nuclear Physics B Supplement}

\end{quotation}
\vskip .2cm
December 1995\\
\end{titlepage}
\vfill
\eject

\def\baselinestretch{1.2}
\baselineskip 16 pt
\noindent
Hyper--Kahler geometry is a much studied subject in modern
theoretical
physics, especially in connection with the theory of gravitational
instantons, supersymmetric models and supergravity, and various
moduli
problems in monopole physics, string theory and elsewhere. The
hyper--Kahler
condition is equivalent to the self--duality (or anti--self--duality)
of a
metric in four dimensions, which is in turn equivalent to Ricci
flatness
plus the Kahler condition. Hyper--Kahler spaces admit three
independent
complex structures $I$, $J$, $K$, satisfying the quaternion algebra
identities,
\begin{equation}
I^2 = J^2 =K^2 = -1 ~ , ~~~~ IJ = - JI = K ~ , ~~~ etc.,
\end{equation}
and hence there is a whole sphere of complex structures generated by
$\alpha I + \beta J + \gamma K$, provided that
${\alpha}^2 + {\beta}^2 + {\gamma}^2 = 1$. Because of this property,
twistor
theory provides a natural framework for studying the algebraic
properties
and the explicit construction of hyper--Kahler metrics [1].
In this paper we study
metrics with isometries that do not preserve the complex structures,
in that
the action of the corrresponding Killing vector fields on the complex
structures
is non--tri--holomorphic; such isometries are also known in the
literature
as rotational.
Our main interest in this subject arises from string theory, where
T-duality
transformations with respect to rotational isometries lead to
non--local
realizations of the $N=4$ world--sheet supersymmetry using
parafermion--like
variables in the dual formulation of certain
superstring vacua [2]. Instead of adopting the
mini--twistor approach for hyper--Kahler metrics with isometries, we
will
use in our discussion a local description in terms of adapted
coordinates,
where the rotational isometry is manifest, and the self--duality
condition
on the metric becomes a non--linear differential equation in the
three (reduced)
dimensions, known as the continual Toda equation.

We focus on the special class of 4--dim hyper--Kahler metrics with
$SO(3)$
symmetry, whose line element in the Bianchi IX formalism is
\begin{equation}
ds^2 = f^2 (t) dt^2 + a^2 (t) {\sigma}_{1}^2 + b^2 (t) {\sigma}_{2}^2
+
c^2 (t) {\sigma}_{3}^2 .
\end{equation}
Here, ${\sigma}_{1}$, ${\sigma}_{2}$, ${\sigma}_{3}$ are the
invariant 1--forms
of $SO(3)$,
\ba
{\sigma}_{1} & = & {1 \over 2} (\sin \psi d \theta - \sin \theta \cos
\psi
d \phi ) , \nonumber\\
{\sigma}_{2} & = & -{1 \over 2} (\cos \psi d \theta + \sin \theta
\sin \psi
d \phi ) , \nonumber\\
{\sigma}_{3} & = & {1 \over 2} (d \psi + \cos \theta d \phi ) ,
\ea
where $\theta$, $\psi$, $\phi$ are the Euler angles, and the
normalization has
been chosen so that ${\sigma}_i \wedge {\sigma}_j = {1 \over 2}
{\epsilon}_{ijk}
d {\sigma}_k$. The coordinate $t$ of the metric can always be chosen
so that
\begin{equation}
f(t) = {1 \over 2} a b c ,
\end{equation}
using a suitable reparametrization. We already see that
$\partial / \partial \phi$ is a manifest Killing vector field. The
Killing
coordinates associated with the other two generators of $SO(3)$ can
not be made
simultaneously manifest, since otherwise they would commute with
$\partial / \partial \phi$, contrary to the
non--abelian nature of the $SO(3)$ isometry group. It may also happen
that
specific solutions of the form (2) exhibit additional isometries,
which is
indeed the case as we will see in the following.

It was established some time ago [3] that the second--order
differential equations
that provide the self--duality condition for the class of metrics
(2), can be
integrated once to yield the following first--order system in $t$:
\ba
2 {a^{\prime} \over a} & = & b^2 + c^2 - 2 {\lambda}_1 bc
- a^2 , \nonumber\\
2 {b^{\prime} \over b} & = & c^2 + a^2 - 2 {\lambda}_2 ca
- b^2 , \nonumber\\
2 {c^{\prime} \over c} & = & a^2 + b^2 - 2 {\lambda}_{3} ab - c^2 ,
\ea
where the three parameters ${\lambda}_i$ remain undetermined for the
moment.
The derivatives (denoted by prime) are taken with respect to $t$,
which satisfies
the constraint (4), and there is also an overall sign ambiguity in
(5) depending
on the self--dual or the anti--self--dual character of the metric;
the two
cases are related to each other by leting $t \rightarrow -t$. Next we
present the only three solutions that exist in this class
leading to complete, regular,
$SO(3)$--invariant hyper--Kahler 4--metrics, namely Eguchi--Hanson,
Taub--NUT and Atiyah--Hitchin.

(i) \underline{Eguchi--Hanson metric} : This metric corresponds to
${\lambda}_1 = {\lambda}_2 = {\lambda}_3 = 0$, and its standard form
is given by the line element
\begin{equation}
ds_{EH}^2 = {dr^2 \over 1 - \left( {m \over r} \right)^4} + r^2
\left( {\sigma}_{1}^2 + {\sigma}_{2}^2 + \left( 1 - \left( {m \over
r}
\right)^4 \right) {\sigma}_{3}^2 \right) ,
\end{equation}
where $m$ is the moduli parameter (for a review, see for instance
[4]).
For $m = 0$ we obtain the flat space
limit of the metric in a manifest $SO(3)$--invariant notation. The
coordinate $r$ is related to $t$, satisfying the normalization (4),
by
\begin{equation}
r^2 = m^2 \coth (m^2 t) .
\end{equation}

(ii) \underline{Taub--NUT metric} : This metric corresponds to
${\lambda}_1 = {\lambda}_2 = {\lambda}_3 = 1$, and its standard form
is
given by the line element
\begin{equation}
ds_{TN}^2 = {1 \over 4} {r + m \over r - m} dr^2 + (r^2 - m^2)
({\sigma}_{1}^2 + {\sigma}_{2}^2 ) + 4m^2 {r - m \over r + m}
{\sigma}_{3}^2 ,
\end{equation}
where $m$ is the relevant moduli parameter (for a review, see for
instance [4]).
Again the coordinate $r$ does not
satisfy the normalization (4), but it is related to $t$ by
\begin{equation}
r = m + {1 \over 2mt} .
\end{equation}

Note that these two solutions exhibit a bigger group of isometries
$SO(3) \times SO(2)$ because $a^2 = b^2$, and hence $\partial /
\partial \psi$
generates an additional manifest isometry that commutes with the
generators
of $SO(3)$. It was thought until 1985 that these two metrics exhaust
the list
of complete $SO(3)$--invariant hyper--Kahler 4--metrics [3]. However,
it also became clear
at that time that the moduli space $M_{2}^0$ of BPS $SU(2)$ monopoles
of
magnetic charge 2 defines a new hyper--Kahler metric with $SO(3)$
isometry. Geodesics in $M_{2}^0$ describe the motion of slowly moving
monopoles,
and certain physical arguments due to Manton claimed that the
scattering of
such monopoles can generate electric charge,
thus converting monopoles into dyons [5].
On the other hand,
it was known that the metric on the moduli space of well--separated
monopoles
can be approximated by the Taub--NUT limit, which exhibits an
additional $SO(2)$
isometry coming from $a^2 = b^2$.
To prove Manton's conjecture as the two monopoles
come close to each other, one had to find a new solution with all
metric
coefficients $a$, $b$, $c$ unequal, so that the absence of such an
additional
$SO(2)$ isometry could explain the generation of electric charge. The
metric in
question was finally found in closed form by Atiyah and Hitchin [6]
(but see
also [7]), and it was
further shown that together with the Eguchi--Hanson and Taub--NUT
metrics,
these three solutions complete the classification of the regular
hyper--Kahler 4--metrics with $SO(3)$ symmetry [8]. Other solutions
are of
course possible, and they have been constructed in the literature
[3], but they
contain sigularities and hence they are not of interest to us.

(iii) \underline{Atiyah--Hitchin metric} : This metric also has
${\lambda}_1 = {\lambda}_2 = {\lambda}_3 = 1$, as in the case of
Taub--NUT, but
the metric coefficients are not all positive; $a$, $b$ are positive,
while
$c$ is taken negative.
A particularly useful parametrization of the corresponding
line element is [6, 7]
\begin{equation}
ds_{AH}^2 = {1 \over 4} a^2 b^2 c^2 {\left( {dk \over k
{k^{\prime}}^2 K^2}
\right)}^2 + a^2 (k) {\sigma}_{1}^2 + b^2 (k) {\sigma}_{2}^2
+ c^2 (k) {\sigma}_{3}^2 ,
\end{equation}
where $a$, $b$, $c$ are given as functions of $k$,
\ba
ab & = & - K(k) (E(k) - K(k)) , \nonumber\\
bc & = & - K(k) (E(k) - {k^{\prime}}^2 K(k)) , \nonumber\\
ac & = & - K(k) E(k) .
\ea
Here, $K(k)$ and $E(k)$ are the complete elliptic integrals of the
first
and second kind respectively, with $0 < k < 1$, and ${k^{\prime}}^2 =
1 - k^2$ is the complementary modulus. The coordinate $t$ in the
parametrization (4) is given by the change of variables
\begin{equation}
t = - {2 K(k^{\prime}) \over \pi K(k)} ~ ,
\end{equation}
up to an additive constant. The verification of equations (5) follows
from the differential equations obeyed by the complete elliptic
integrals. We also note using standard expansions of the elliptic
integrals that the Atiyah--Hitchin metric approaches the Taub--NUT
limit as
$k \rightarrow 1$, in which case one obtains the Taub--NUT metric
with
negative moduli (mass) parameter $m = - 1/2$.

Having presented the complete list of the $SO(3)$--invariant
hyper--Kahler
4--metrics, we are now in the position to examine their Toda frame
formulation that exploits the presence of rotational isometries and
leads
to the explicit construction of all three complex structures as an
$SO(2)$--doublet
plus a singlet. Recall first the precise definition that
distinguishes
a rotational from a translational (also known as tri--holomorphic)
Killing vector field $K_{\mu}$. If $K_{\mu}$ satisfies the condition
\begin{equation}
{\nabla}_{\nu} K_{\mu} = \pm {1 \over 2} \sqrt{\det g} ~
{{\epsilon}_{\nu \mu}}^{\kappa \lambda} {\nabla}_{\kappa} K_{\lambda}
,
\end{equation}
it will be called translational, otherwise it will be rotational [9,
10]. Here,
the $\pm$ sign is chosen according to the self--dual or the
anti--self--dual
nature of the underlying 4--metric $g_{\mu \nu}$. A more algebraic
description
of the character of a Killing vector field, which will be useful in
the
following, is given in terms of adapted coordinates for the metric
\begin{equation}
ds^2 = V (d \tau + {\omega}_{i} d x^i )^2 + V^{-1} {\gamma}_{ij} dx^i
dx^j ,
\end{equation}
where $V$, ${\omega}_{i}$, ${\gamma}_{ij}$ are all independent of
$\tau$ and
$K = \partial / \partial \tau$, using the notion of the nut
potential. The
nut potential $b_{nut}$ of a generic vacuum metric (14) that
satisfies
Einstein's equations is defined using the isometry
$\partial / \partial \tau$, as follows:
\begin{equation}
{\partial}_{i} b_{nut} = {1 \over 2} V^2 \sqrt{\det \gamma} ~
{{\epsilon}_{i}}^{jk} ({\partial}_{j} {\omega}_{k} - {\partial}_{k}
{\omega}_{j}) .
\end{equation}
For rotational isometries generated by $\partial / \partial \tau$,
the
quantity
\begin{equation}
S_{\pm} = b_{nut} \pm V
\end{equation}
is coordinate dependent (it will be constant only for translational
isometries),
and it can always be chosen equal to one of the coordinates,
say $x^3 = z$, up to an overall
normalization given by $1 / \sqrt{{\gamma}^{ij} ({\partial}_{i}
S_{\pm})
({\partial}_{j} S_{\pm})}$. Then, the other two coordinates $x^1 = x$
and
$x^2 = y$ can be chosen so that without loss of generality the metric
(14) has elements
\begin{equation}
V^{-1} = {\partial}_{z} \Psi ~ , ~~~~ {\omega}_{1} = \mp
{\partial}_{y} \Psi ~ ,
{}~~~~ {\omega}_{2} = \pm {\partial}_{x} \Psi ~ , ~~~~ {\omega}_{3} =
0 ,
\end{equation}
and diagonal $\gamma$--metric
\begin{equation}
{\gamma}_{11} = {\gamma}_{22} = e^{\Psi} ~ , ~~~~ {\gamma}_{33} = 1 ,
\end{equation}
all of which are determined in terms of a
single scalar function $\Psi (x, y, z)$ [9, 10]. With this
choice, which is possible only for hyper--Kahler metrics, the
function $\Psi$
satisfies the so--called 3--dim continual Toda equation
\begin{equation}
({\partial}_{x}^2 + {\partial}_{y}^2 ) \Psi + {\partial}_{z}^2
(e^{\Psi}) = 0 .
\end{equation}
The continual Toda equation arises as a large $N$ limit of the
ordinary 2--dim
Toda theory based on the group $SU(N)$, where the Dynkin diagram
becomes a
continuous line parametrized by the third space variable $z$ [11,
12]. In the
context of general relativity, this equation first appeared in the
work of
Boyer and Finley [9].

Our aim is to examine the rotational versus the translational
character of the
Killing vector fields of the above three $SO(3)$--invariant
hyper--Kahler
4--metrics [8], and use this formalism to construct explicitly the
corresponding
Toda potentials of the metrics. The common element of these three
metrics is
the presence of (at least one) rotational isometry, and this is where
our
analysis will come into play. In particular, one finds that the
$SO(3)$
generators of the Eguchi--Hanson metric are translational, while the
additional
$SO(2)$ isometry $\partial / \partial \psi$ coming from $a^2 = b^2$
is
rotational. For the Taub--NUT
metric the situation is reversed with the $SO(3)$ generators acting
as
rotational isometries and $\partial / \partial \psi$ being
translational.
Finally, for the Atiyah--Hitchin metric the generators of the $SO(3)$
isometry
are rotational; in fact this solution provides the only example
known to this date of a complete
hyper--Kahler 4--metric that is purely rotational without
exhibiting any translational isometries. We now list the relevant
coordinate transformations that bring these three metrics into their
respective
Toda frames, and choose our conventions so that the relevant object
to
consider is $S_{+}$ instead of $S_{-}$. Further details can be
found in the report of our recent work [13].

(i) \underline{Eguchi--Hanson metric} : Using the rotational Killing
vector
field $\partial / \partial \psi$, we write the metric (6) in the form
(14)
and determine the explicit expression for $V$ and $b_{nut}$.
This allows to find
the coordinate $z$ according to the general theory we have outlined.
The
complete transformation to the Toda frame (17)--(18) can be easily
performed
thanks to the diagonal form of the metric. The result is summarized
as follows:
\ba
x & = & 2 \sqrt{2} \cos \phi \tan {\theta \over 2} ~ , ~~~~
y = 2 \sqrt{2} \sin \phi \tan {\theta \over 2} , \nonumber\\
z & = & {1 \over 4} r^2 ~ , ~~~~ \tau = 2(\psi + \phi) .
\ea
Then, the corresponding Toda potential turns out to be
\begin{equation}
e^{\Psi} = {z^2 - {\alpha}^2 \over 2 \left(1 + {1 \over 8} (x^2 +
y^2)
\right)^2} , ~~~~ z^2 \geq {\alpha}^2 ,
\end{equation}
where $4 \alpha = m^2$, and this is clearly a solution of the
continual
Toda equation (19). The presence of the other isometry
$\partial / \partial \phi$ reflects into the symmetric quadratic
dependence
of $\Psi$ on $x$ and $y$.

(ii) \underline{Taub--NUT metric} : For the Taub--NUT metric we may
use
$\partial / \partial \phi$ as the generator of a rotational isometry,
and
after some calculation we find that the change of coordinates that
brings
the line element (8) into a manifest Toda frame form is given by
\ba
x & = & \psi , ~~~~ y = - {1 \over 4 m^2 t} \cos \theta +
\log \left( \tan {\theta \over 2} \right) , \nonumber\\
z & = & {1 \over 4t} \left(1 + {1 \over 8 m^2 t} {\sin}^2 \theta
\right) ,
{}~~~~ \tau = 2 \phi .
\ea
Here, the coordinate $t$ is related to the standard variable $r$ by
equation (9). The Toda potential of this metric turns out to be
\begin{equation}
e^{\Psi} = {1 \over 16 t^2} {\sin}^2 \theta .
\end{equation}
We remark that the presence of the additional $SO(2)$ isometry
$\partial / \partial \psi$ reflects into the form of $\Psi$ as
$x$--independence. This Toda potential is also a solution of the
non--linear equation (19), as it is required on general grounds,
but it can not be expressed as an explicit
function of $y$ and $z$ in closed form.

(iii) \underline{Atiyah--Hitchin metric} : Like the Taub--NUT metric,
we
also use here the rotational character of the Killing vector field
$\partial / \partial \phi$ in order to perform the transformation to
a Toda frame. The result, which is much more complicated in
comparison
to the Eguchi--Hanson and Taub--NUT metrics, was first presented in
the
literature by Olivier [14]. Using our notation, and introducing the
complex
variable
\begin{equation}
\nu = \log \left( \tan {\theta \over 2} \right) + i \psi ,
\end{equation}
the relevant transformation to the Toda frame reads:
\begin{equation}
y + ix = K(k) \sqrt{1 + {k^{\prime}}^2 {\sinh}^2 \nu}
\left( \cos \theta + {\tanh \nu \over K(k)}
\int_{0}^{\pi / 2} d \gamma {\sqrt{1 - k^2 {\sin}^2 \gamma} \over
1 - k^2 {\tanh}^2 \nu {\sin}^2 \gamma} \right) ,
\end{equation}
\begin{equation}
z = {1 \over 8} K^2 (k) \left( k^2 {\sin}^2 \theta + {k^{\prime}}^2
(1 + {\sin}^2 \theta {\sin}^2 \psi ) - 2 {E(k) \over K(k)} \right) ,
\end{equation}
and
\begin{equation}
\tau = 2 \left( \phi + \arg (1 + {k^{\prime}}^2 {\sinh}^2 \nu)
\right) .
\end{equation}
We also obtain the Toda potential of the Atiyah--Hitchin metric,
\begin{equation}
e^{\Psi} = {1 \over 16} K^2 (k) {\sin}^2 \theta \mid 1 +
{k^{\prime}}^2
{\sinh}^2 \nu \mid .
\end{equation}
It can be easily verified that in the limit $k \rightarrow 1$ the
above
expressions yield the corresponding results for the Taub--NUT metric.
It is also impossible here to write $\Psi$ as an explicit function of
$x$, $y$ and $z$ in closed form.

We now turn to the problem of the explicit consruction of all three
complex structures, using the Toda frame formulation. In a
system like the Toda frame, which uses local coordinates that are
adapted to a rotational isometry, the corresponding three Kahler
forms
are not all invariant under $\tau$--shifts. This is another way of
saying that
rotational isometries have a non--triholomorphic action. In fact, the
Kahler forms in this case group into an $SO(2)$--doublet plus a
singlet,
which can be written down explicitly [2] as follows: The doublet is
\begin{equation}
\left(\begin{array}{c}
F_{1} \\
      \\
F_{2} \end{array} \right) = e^{{1 \over 2} \Psi}
\left(\begin{array}{cr}
\cos {\tau \over 2} & \sin {\tau \over 2} \\
                    &                     \\
\sin {\tau \over 2} & - \cos {\tau \over 2} \end{array} \right)
\left(\begin{array}{c}
f_{1} \\
      \\
f_{2} \end{array} \right)
\end{equation}
where
\ba
f_{1} & = & (d \tau + {\omega}_{2} dy) \wedge dx - V^{-1}
dz \wedge dy ~ , \nonumber \\
f_{2} & = & (d \tau + {\omega}_{1} dx) \wedge dy + V^{-1} dz \wedge
dx ~ ,
\ea
while the singlet is
\begin{equation}
F_{3} = (d \tau + {\omega}_{1} dx + {\omega}_{2} dy) \wedge dz +
V^{-1} e^{\Psi} dx \wedge dy ~ .
\end{equation}
If we were considering the action of the whole $SO(3)$ isometry on
the
three Kahler forms, it would turn out that either they form a triplet
when all the generators
are rotational, as in Taub--NUT and Atiyah--Hitchin metrics, or they
are
all singlets when the generators are
translational, as in the Eguchi--Hanson metric [8].

The complex structures of the Eguchi--Hanson and Taub--NUT metrics
are
already known in the literature, because these metrics exhibit (at
least) one
translational isometry, and there are simple formulas for them in
adapted translational coordinates (see for instance [8]).
Nevertheless, we may use the above
result to find the form of the three complex structures in the
Toda frame formulation of these two metrics as well.
However, for the Atiyah--Hitchin
metric, which is purely rotational, the use of our result provides 
a way to
perform the explicit construction of all three Kahler forms. It is
straightforward, but a very tedious exercise, to use the
transformations
(25)--(28) and obtain the result for $F_1$, $F_2$, $F_3$ directly in
terms
of the original variables $\theta$, $\phi$, $\psi$ and $k$ of the
Atiyah--Hitchin metric. 
There is also an alternative explicit construction of the complex
structures as an $SO(3)$--triplet.
We hope to make available the details of
these
expressions in another publication [15].

In conclusion, note that our study of rotational hyper--Kahler
geometry has
so far been limited to the class of $SO(3)$--invariant metrics. The
Atiyah--Hitchin metric is very special, exhibiting only rotational
isometries,
and it can be regarded as the simplest representative from a class of
purely rotational metrics. The construction of a descending series of
new
4--dim $SO(2)$--invariant hyper--Kahler metrics, if they indeed
exist,
will certainly require a deeper understanding of the special features
of
the Atiyah--Hitchin metric, using many different points of view. One
way
to understand the qualitative differences of the Atiyah--Hitchin
metric
from the other two solutions is provided by the free field
realization of
the corresponding Toda potentials. An investigation along these lines
will
be reported elsewhere [13]. We only mention here that a drawback of
our
method is the local nature of the Toda construction; although every
solution
of the continual Toda equation corresponds to a 4--dim hyper--Kahler
metric,
the issue of their completeness requires a separate and more delicate
study.

\vskip .3in
\centerline{\bf Acknowledgements}
\noindent
One of us (I.B.) wishes to thank the organizers of the Buckow
Symposium for
their kind invitation and financial support that made possible his
participation to this very enjoyable event.

\newpage

\centerline{\bf REFERENCES}
\begin{enumerate}
\item N. Hitchin, A. Karlhede, U. Lindstrom and M. Rocek, Comm. Math.
Phys.
\underline{108} (1987) 535.
\item I. Bakas and K. Sfetsos, Phys. Lett. \underline{B349} (1995)
448.
\item G. Gibbons and C. Pope, Comm. Math. Phys. \underline{66} (1979)
267.
\item T. Eguchi, P. Gilkey and A. Hanson, Phys. Rep. \underline{66}
(1980)
213.
\item N. Manton, Phys. Lett. \underline{B110} (1982) 54; in
``Monopoles in
Quantum Field Theory", World Scientific, Singapore, 1981.
\item M. Atiyah and N. Hitchin, Phys. Lett. \underline{A107} (1985)
21;
Phil. Trans. R. Soc. Lond. \underline{A315} (1985) 459; ``The
Geometry
and Dynamics of Magnetic Monopoles", Princeton University Press,
Princeton,
1988.
\item G. Gibbons and N. Manton, Nucl. Phys. \underline{B274} (1986)
183.
\item G. Gibbons and P. Rubback, Comm. Math. Phys. \underline{115}
(1988)
267.
\item C. Boyer and J. Finley, J. Math. Phys. \underline{23} (1982)
1126.
\item J. Gegenberg and A. Das, Gen. Rel. Grav. \underline{16} (1984)
817.
\item M. Saveliev, Comm. Math. Phys. \underline{121} (1989) 283;
Theor.
Math. Phys. \underline{92} (1993) 1024.
\item I. Bakas, in ``Supermembranes and Physics in 2+1 Dimensions",
eds.
M. Duff, C. Pope and E. Sezgin, World Scientific, Singapore, 1990;
Comm. Math. Phys. \underline{134} (1990) 487; Q.--H. Park, Phys.
Lett.
\underline{B238} (1990) 287.
\item I. Bakas and K. Sfetsos, ``Toda Fields of $SO(3)$ Hyper--Kahler
Metrics
and Free Field Realizations", CERN/ENSLAPP/Utrecht preprint (to
appear).
\item D. Olivier, Gen. Rel. Grav. \underline{23} (1991) 1349.
\item K. Sfetsos, Utrecht preprint (to appear).
\end{enumerate}

\end{document}